\documentclass[aps,pre,onecolumn]{revtex4}
\usepackage{amsfonts,amssymb,amsmath,bm}
\usepackage[dvips]{graphicx}
\usepackage{color}
\usepackage[T1,T2A]{fontenc}
\usepackage[cp1251]{inputenc}
\usepackage[english,russian]{babel}

\begin{document}

\title{Emergence of Nanoscale Modulation in Liquid Crystals as a Result of Short-Range Order Parameter Condensation}

\author{E.I.Kats}
\affiliation {Laue-Langevin Institute, Grenoble, France (Retired), and \\
Landau Institute for Theoretical Physics, RAS, \\ 
Chernogolovka, Moscow region, Russia (Retired)}

\begin{abstract}
The discovery of the twist-bend nematic phase in liquid crystals composed of bent-core and dimeric molecules has revealed an unexpected mechanism for the spontaneous formation of nanoscale periodic structures in soft condensed matter.
Unlike conventional liquid-crystalline phases, the twist-bend phase ($N_{TB}$) exhibits a nanoscale heliconical modulation despite the absence of molecular chirality. In this note, dedicated to the late R. Meyer, we discuss a Landau phenomenological interpretation of this phenomenon. The central idea is that a short-range orientational order parameter undergoes condensation, giving rise to a heliconical structure characterized by a finite wave vector. We emphasize that the conventional nematic order is already long-ranged, whereas the additional order parameter describes local orientational correlations hidden within the nematic state. The resulting phase transition bears a close analogy to the de Gennes theory of the nematic--smectic-A transition. Fluctuation effects are expected to drive the transition weakly first order. The theory also predicts a new Goldstone mode associated with the spontaneously broken continuous symmetry of the heliconical state

\end{abstract}

\maketitle

{\bf {Preface.}}${\, }$

This note is dedicated to the memory of the late R. Meyer, whose pioneering ideas continue to inspire research in liquid-crystal physics. 
Although I never had the opportunity to know R. Meyer personally, I have long admired his work and its profound scientific significance. In particular, I have always been impressed by his classic and widely cited contribution \cite{Meyer}, especially considering that it appeared in a conference volume that is not readily available in most libraries.

The spontaneous emergence of the heliconical state in the $N_{TB}$ phase may be viewed as a realization of ideas originating in Meyer's pioneering work on spontaneous bend deformations and flexoelectric coupling. Truly outstanding scientific contributions, such as Meyer's work \cite{Meyer}, continue to attract increasing attention and stimulate new developments long after their publication.

The purpose of this note is to emphasize that the softening of the Frank elastic modulus—even if it were to approach zero—is not the fundamental mechanism driving the $N$--$N_{TB}$ phase transition. Although the Frank modulus does soften in the vicinity of the transition, this softening is itself a consequence of the coupling between the conventional long-wavelength nematic order parameter and a new short-wavelength order parameter describing local orientational correlations hidden within the nematic state. It is the condensation of this short-range order parameter, rather than the softening of the Frank elasticity itself, that constitutes the primary physical mechanism responsible for the transition.

{\bf {Introduction.}}${\, }$

The study of modulated structures occupies a central place in the physics of liquid crystals. Many of the concepts that have shaped this field originated in the pioneering work of R. Meyer, whose insights into spontaneous distortions, flexoelectricity, and chiral liquid-crystalline order profoundly influenced our understanding of soft condensed matter.

The discovery of the twist-bend nematic ($N_{TB}$) phase opened a new chapter in this field. In this phase, the molecular orientation adopts a heliconical structure with a nanoscale pitch, typically only a few molecular lengths. The phenomenon is particularly remarkable because it occurs in systems composed of achiral molecules. Consequently, the emergence of the heliconical structure represents a striking example of spontaneous chiral symmetry breaking.

A particularly intriguing aspect of the transition is that it occurs between two phases that are both already ordered. The high-temperature nematic phase possesses long-range orientational order described by the director field ${\bf n}$. Nevertheless, experiments indicate the presence of additional local orientational correlations associated with the tendency of bent-core or dimeric molecules to develop local bend and polar order. In the nematic phase, however, these correlations remain short-ranged and fluctuate in space, producing no additional macroscopic symmetry breaking beyond that already present in the nematic state.

The central idea of this note is that the twist-bend transition is driven not by the conventional nematic order parameter, which remains finite on both sides of the transition, but by the condensation of a secondary order parameter describing these local orientational correlations. As the temperature decreases, fluctuations of this hidden order parameter become critical at a finite wave vector and develop long-range coherence. The resulting ordered state is a periodically modulated heliconical structure with a nanoscale pitch.

From this perspective, the twist-bend phase provides an example of ordering within an already ordered medium. Long-range nematic order is present throughout the transition, while a previously hidden short-range correlated mode undergoes condensation, giving rise to a new level of structural organization.

A related example of sequential symmetry breaking is provided by
superfluid $He^{3}$ \cite{MD19}  confined in a nematically ordered
aerogel. In this system, cooling from the normal state produces a
sequence of the polar, polar-distorted $PdA$, and polar-distorted
$PdB$ phases. The polar phase possesses a nematic-like orbital
anisotropy, while at the subsequent transition to the $PdA$ phase an
additional orbital component develops, introducing a second vector
orthogonal to the polar anisotropy axis. The resulting orbital triad,
gives rise to a chiral state. This system provides a useful example in
which chirality emerges at a secondary symmetry-breaking transition
from a pre-existing nematic-like ordered state. It is worth also to note \cite{KL82}, where the secondary symmetry breaking phase transition leads to the formation of a special type of composite defects, such as walls bounded by strings.

{\bf {Landau description of the heliconical instability.}}${\, }$

As discussed above, the spontaneous emergence of a structure modulated by a finite wave vector $q_0$ is characterized by the appearance of a short-wavelength order parameter, denoted hereafter by $\bm{\varphi}$. Within the framework of Landau theory, the fundamental distinction between the conventional long-wavelength order parameter, already present in the nematic phase and described by the director field $\mathbf{n}$ or the quadrupolar tensor $Q_{ik}$, and the short-wavelength order parameter $\bm{\varphi}$ describing the $N_{TB}$ phase lies in the form of the corresponding gradient terms in the free-energy expansion. Fluctuations of the order parameter play an essential role in the phase transition provided that their excitation energy becomes sufficiently small; that is, the fluctuations become soft.

The softness of these fluctuations is determined by the gradient contribution $F_g$ to the free energy. For a long-wavelength order parameter, the gradient energy has the familiar form
\begin{equation}
\label{k1}
F^{lw}_g \propto (\nabla {\hat{Q}})^2 .
\end{equation}

From the form (\ref{k1}), it can be inferred that the softening of fluctuations in this case occurs uniformly in space, or in the vicinity of an isolated point $q=0$ in reciprocal (Fourier) space.

Conversely, short-wavelength modulation of the order parameter implies that fluctuations become soft (their excitation energy is low) 
in the vicinity of a certain manifold in Fourier space. Depending on the system's symmetry, this manifold may consist of several points 
${\bf q} = (\pm q_{0x} , \pm q_{0y} , \pm q_{0z})$
or a neighborhood of a circle $ |{\bf q}| = q_{0 \perp}$, and $q_{0z} = 0$), or a sphere $|{\bf q}| = q_0$. The latter manifolds
can be either ideal or corrugated.

To take a concrete example of the $N_{TB}$ liquid crystals, with a phase transition into a short-wavelength (the order parameter $\bm \varphi $) modulated structure arising 
in the background of a developed long-wave orientational order parameter (director $\bf n$) of the nematic $N$, the corresponding 
(short-wavelength) gradient energy can be represented in the following form

\begin{eqnarray}
F^{sw}_g \propto
 \left[
 \left(n_i n_k \partial_i \partial_k +q_0^2\right) \bm\varphi \right]^2
.
\label{k2}
 \end{eqnarray}
To describe the $N$--$N_{TB}$ phase transition, the gradient term (\ref{k2}) must be supplemented by the usual local terms of the Landau expansion together with coupling terms between the long-wavelength fluctuations of the director field $\mathbf n$ and the short-wavelength fluctuations of $\bm{\varphi}$. This analysis was carried out in Ref.~\cite{KatsLebedev}; here we summarize only those results that are essential for the present discussion.

A distinctive feature of the $N$--$N_{TB}$ transition is the coexistence of two qualitatively different order parameters. The first is the conventional nematic order parameter represented by the director field $\mathbf n$. Its ordering occurs at wave vector $q=0$ and therefore corresponds to a spatially uniform broken-symmetry state. This long-range orientational order is established at the isotropic--nematic transition and persists throughout both the nematic and twist-bend phases.

The second order parameter, $\bm{\varphi}$, characterizes local orientational correlations within the nematic phase. Physically, these correlations arise from the tendency of bent-core or dimeric molecules to develop local polar and bend order \cite{BK13}--\cite{CD11}. In the nematic phase, however, this order remains short-ranged: its spatial average vanishes, while its correlation function decays over a finite characteristic length.

The central assumption of the Landau theory is that these short-range correlations become critical upon cooling. Unlike the conventional nematic order parameter, whose instability would occur at $q=0$, the susceptibility associated with $\bm{\varphi}$ reaches its maximum at a finite wave vector $q_0$. Consequently, the soft mode that condenses carries a nonzero wave vector. The transition therefore resembles a generalized crystallization process in which long-range order first develops at a finite spatial periodicity rather than in a spatially uniform state.

This distinction between condensation at $q=0$ and at $q=q_0$ is fundamental. In conventional continuous phase transitions, such as the isotropic--nematic transition, critical fluctuations occur at zero wave vector and produce a homogeneous ordered phase. In contrast, condensation 
at finite $q_0$ necessarily gives rise to a periodically modulated structure with characteristic period
$$
\Lambda=\frac{2\pi}{q_0}.
$$
To isolate the short-wavelength component of the order parameter, we write $\bm{\varphi}$ in terms of a slowly varying complex field $\bm{\psi}$,
 \begin{equation}
 \bm\varphi= 2\,\mathrm{Re}\
 \left[\bm\psi \exp(i q_0 z) \right] ,
 \label{bana3}
 \end{equation}
where the preferred direction of the nematic director is taken to be along the $z$ axis. Unlike $\bm{\varphi}$, the complex field $\bm{\psi}$ varies only on long length scales and is perpendicular to the $z$ axis,
$\bm{\psi}=(\psi_x,\psi_y,0)$.

Minimization of the Landau free energy with respect to $\bm{\psi}$ yields the solutions $\psi_x=i\psi_y$ or $\psi_x=-i\psi_y$, for which $\bm{\psi}^2=0$. These correspond to the heliconical structure
 \begin{equation}
 \varphi_x=2 |\psi_x|\cos(q_0z+\phi), \quad
 \varphi_y=\pm 2|\psi_x|\sin(q_0z+\phi),
 \label{cone}
 \end{equation}
where $\phi$ denotes the phase of $\psi_x$, while the two signs correspond to the two possible handednesses of the heliconical structure.

The $N_{TB}$ phase possesses an additional soft (Goldstone) mode associated with slow spatial variations of the phase $\phi$. Because the bulk free energy is invariant under a uniform phase shift, the corresponding elastic energy depends only on gradients of $\phi$. Since the phase is a long-wavelength degree of freedom, its gradient energy has the same general form as Eq.~(\ref{k1}). However, because the $N_{TB}$ phase develops on the background of an anisotropic nematic, the corresponding elastic constants are themselves anisotropic.

The scenario outlined above demonstrates that the $N$--$N_{TB}$ transition does not arise from a loss of stability of the conventional nematic order parameter. The nematic phase already possesses long-range orientational order described by the director field $\mathbf n$, whose expectation value remains finite throughout the transition. Consequently, unlike the isotropic--nematic transition, the emergence of the heliconical phase is not associated with the onset of orientational order from an initially disordered state.

Instead, the transition is driven by a second order parameter describing local orientational correlations. Upon cooling, fluctuations of this hidden order parameter become increasingly pronounced until they eventually develop long-range coherence. From this perspective, the twist-bend phase emerges not because the nematic order parameter becomes unstable, but because a previously hidden short-range mode embedded within the nematic state undergoes condensation. The resulting heliconical structure therefore represents a new level of ordering superimposed on an already ordered nematic background.
where $\phi$ is the phase of $\psi_x$ and signs $\pm$ correspond to two possible rotation directions of the conical structure.
It is worth noting one additional soft (Goldstone) mode in the $N_{TB}$ phase related to long-scale variations of the phase $\phi$ in the expression (\ref{cone}). Since the bulk energy is independent of a homogeneous phase shift, the elastic energy related to variations of $\phi$ depends only on its gradient. 
Since the phase of the order parameter is a long-wavelength degree of freedom (i.e., defined at scales larger than the correlation length), the corresponding gradient energy has the standard 
form (\ref{k1}). However, because the $N_{TB}$ state arises on the background of a preexisting anisotropic nematic, this gradient energy must also be anisotropic.

{\bf {Analogy with De Gennes' theory of the nematic-smectic $A$ phase transition.}}${\, }$

An instructive analogy exists between the twist-bend transition and de Gennes' theory of the nematic--smectic-$A$ phase transition. The principal difference lies in the nature of the order parameter: in the present case it is the two-component vector $\bm{\psi}$, whereas in the de Gennes theory the order parameter is a complex scalar field. We do not review this elegant theory here, since detailed accounts may be found in many standard textbooks on liquid-crystal physics (see, e.g., \cite{GP93}--\cite{BL11}).

In the smectic-$A$ case, the critical order parameter describes a density wave with a finite wave vector. Mean-field theory predicts a continuous phase transition. However, de Gennes demonstrated that the smectic order parameter couples strongly to fluctuations of the nematic director. This coupling generates singular fluctuation corrections that qualitatively modify the critical behavior.

A closely related mechanism operates in the twist-bend transition. The order parameter responsible for heliconical ordering is coupled to the Goldstone modes associated with the nematic orientational order. Within mean-field theory the transition is continuous and belongs to a conventional universality class. Once fluctuations are taken into account, however, the close analogy with the de Gennes problem becomes apparent.

The fluctuation-induced coupling produces nonanalytic contributions to the free energy and destabilizes the continuous transition, rendering it weakly first order. This prediction provides a natural explanation for the experimentally observed behavior of many twist-bend materials, where small discontinuities and phase coexistence are commonly observed. Thus, the $N$--$N_{TB}$ transition provides another important example of a fluctuation-induced first-order phase transition in soft condensed matter.

{\bf {Chiral $N_{TB}$ liquid crystals.}}${\, }$

Thus far we have considered only the phase transition from an achiral nematic phase $N$, composed of achiral molecular building blocks, to a heliconical tilted phase that is structurally chiral. This transition therefore represents a case of spontaneous chiral symmetry breaking. In the absence of any intrinsic chirality, one expects right-handed and left-handed heliconical domains to occur with equal probability.

A different situation arises for the phase transition from an intrinsically chiral cholesteric phase $Ch$, composed of chiral molecular building blocks, to the chiral $N_{TB}$ phase. In this case the degeneracy between the two helicities is lifted, and the probabilities of forming right-handed and left-handed domains are no longer equal.

The existence of two well-separated characteristic wave vectors in chiral $N_{TB}$ liquid crystals—the cholesteric wave vector $q_s \ll q_0$, where $q_0$ is the wave vector of the heliconical modulation—suggests that this doubly periodic system may be described theoretically using an approach analogous to the theory of the Kapitza pendulum. 

The Kapitza pendulum is a classic example of dynamic stabilization of an inverted rigid pendulum whose pivot undergoes rapid vertical oscillations. The phenomenon was first discovered by Stephenson \cite{ST08} and later independently analyzed by Kapitza \cite{KA51}. Stabilization occurs when the natural frequency of the pendulum, $\omega_s$, is much smaller than the driving frequency, $\omega_0$, namely $\omega_s \ll \omega_0$.

Although the Kapitza pendulum is a dynamical system, whereas the $Ch$--$N_{TB}$ transition is a thermodynamic phenomenon, the underlying physics and the mathematical description are remarkably similar. In both cases, the essential procedure is averaging over rapidly varying degrees of freedom. As a consequence, a critical condition emerges at which the topology of the effective free-energy landscape (for liquid crystals) or the effective Lagrangian (for the mechanical pendulum) changes from a single-well to a double-well structure.

The separation of the two characteristic length scales suggests that the short-period heliconical modulation can be averaged out, leading to an effective theory for the slowly varying cholesteric structure. This procedure is closely analogous to the averaging employed in the theory of the Kapitza pendulum.
Averaging over the rapidly varying modulation described by $\bm{\varphi}$ yields an effective interaction that depends on the relative phase shift $\delta$ between the two helices: the cholesteric helix and the tilted heliconical $N_{TB}$ structure. A simple phenomenological expression for the effective potential is
\begin{eqnarray} 
 V_{eff}(\delta ) = - \alpha \cos (\delta ) + \beta \cos (2 \delta )
, \quad
 \label{kp2}
 \end{eqnarray}
where the phenomenological coefficients $\alpha$ and $\beta$ are functions of the microscopic parameters characterizing the $Ch$--$N_{TB}$ phase transition,
\begin{equation} 
(\alpha , \beta ) = (\alpha , \beta )\left(\frac{q_0}{q_s} , A, g , K\right)\, .
\label{kp3}
\end{equation}
Here $A$ is the amplitude of the twist-bend modulation, $K$ is the Frank elastic constant, and $g$ characterizes the coupling between the director field ${\bf n}$ and the order parameter $\bm{\varphi}$.
The effective potential (\ref{kp2}) immediately shows that, for $\beta<\alpha/4$, the free energy possesses a single minimum at $\delta=0$, corresponding to a co-rotating configuration of the cholesteric and heliconical twists. When $\beta>\alpha/4$, the free energy develops two minima, at $\delta=0$ and $\delta=\pi$. The second minimum corresponds to counter-rotating helices. The critical condition for this topological transition can be estimated in terms of the microscopic parameters as
\begin{eqnarray} 
g^2 A^2 \simeq K q_0^2
. \quad
 \label{kp4}
 \end{eqnarray}

{\bf {Concluding remarks.}}${\, }$

The twist-bend transition is unusual because it occurs between two already ordered phases. The conventional nematic phase is characterized by long-range orientational order associated with the zero wave vector ($q=0$), whereas the twist-bend phase arises from the condensation of an additional order parameter at a finite wave vector. The observed nanoscale modulation is therefore a direct manifestation of finite-wave-vector ordering developing within an already ordered medium. In this sense, the transition provides a striking example of ``ordering within an ordered state.''

The free-energy minimum is reached not for a spatially uniform configuration but for a periodically modulated one. The ordered phase is characterized by a director that precesses around a fixed axis while maintaining a constant tilt angle, thereby forming a heliconical structure. This state simultaneously breaks rotational symmetry and continuous translational symmetry along the helix axis. The modulation period is determined by molecular-scale interactions and is therefore much shorter than optical wavelengths. As a result, the phase remains optically similar to a conventional nematic despite possessing a highly nontrivial internal structure. Most importantly, the modulation is neither externally imposed nor a consequence of molecular chirality. Instead, it emerges spontaneously through the condensation of a previously fluctuating short-range order parameter.

The heliconical phase possesses broken continuous symmetries and therefore supports low-energy collective excitations. In addition to the conventional orientational fluctuations, the modulated structure gives rise to an additional soft mode associated with phase shifts of the helix. Physically, this excitation corresponds to slow displacements of the heliconical pattern along its axis. Because the free energy is invariant under a uniform phase shift, this excitation is a Goldstone mode.

At long wavelengths, the elastic properties of this mode resemble those of layered systems, despite the absence of true density modulation. Thus, the heliconical phase combines characteristic features of nematic and smectic ordering in an unusual way. Experimental observation of this Goldstone mode would provide an important test of the theoretical picture, since its existence is a direct consequence of the spontaneous formation of a periodically modulated orientational state.

The resulting heliconical phase provides a striking example of spontaneous chiral symmetry breaking and self-organization on molecular length scales. The analogy with the de Gennes theory further demonstrates that fluctuations play a decisive role, converting the mean-field continuous transition into a weakly first-order one, in agreement with experimental observations.

Beyond its specific application to twist-bend nematics, the present framework illustrates a more general principle: local correlations hidden within an apparently uniform phase may condense into long-range modulated order when the corresponding collective mode becomes soft. From this perspective, the twist-bend phase represents a paradigmatic example of how nanoscale structure can emerge through the collective organization of short-range order.

The principal conceptual conclusions of this note may be summarized as follows:
\begin{itemize}
\item
The nematic director represents a long-range order parameter condensed at $q=0$.
\item
Hidden polar/bend correlations constitute a short-range order parameter within the nematic phase.
\item
The $N$--$N_{TB}$ transition corresponds to the condensation of this hidden order parameter at a finite wave vector $q_0$.
\item
The nanoscale heliconical structure is a direct consequence of finite-wave-vector condensation.
\end{itemize}
Note to the point that the notion of a secondary orientational order parameter coexisting with an already established nematic has a distinguished precedent. Kamien \cite{KA1} analysed an achiral liquid crystal possessing both nematic and hexatic bond order and showed that coupling of the bond-angle field to the director produces a conical ground state, the analogue of a cholesteric unwound by an applied field, whenever the Frank constants satisfy $K_2 > K_3$. Kamien and Levine \cite{KA2} subsequently demonstrated that boundaries, to which long-ranged orientational correlations are acutely sensitive, can pin the twisting bond order and lock the system into an apparently uniform state. These works anticipate the viewpoint adopted in the present note: an apparently uniform nematic may conceal a short-wavelength order parameter, and it is the condensation of such a parameter at a finite wave vector, rather than the softening of the Frank elasticity itself, that generates the heliconical ground state of the twist-bend phase.

The ideas developed here resonate strongly with the scientific legacy of R. Meyer, whose pioneering work established many of the concepts underlying our present understanding of spontaneous symmetry breaking, elastic instabilities, and the remarkable diversity of orientationally ordered structures in liquid crystals.

\end{document}